\documentclass[12pt]{article}
\author{{\Large\bf  Sandra Carillo}\\[12pt] 
Dipartimento di Scienze di Base e Applicate
per l'Ingegneria  \\
 Universit\`a di Roma {\it La Sapienza}, % \filler \\
Via A.\ Scarpa 16, Rome, Italy   \\[6pt]  
I.N.F.N. - Sezione Roma1, Gr. IV - M.M.N.L.P.,  Rome, Italy  }

\title{\bf A 3-dimensional singular kernel problem in  viscoelasticity:  an existence result}
\usepackage{fancyhdr}
\usepackage{filecontents} 
                     \usepackage{amsmath,amsfonts,amssymb,amscd,latexsym,lineno,epsfig}
\date{} % no date appears 
\newcommand{\N}{{\mathbb N}}
\newcommand{\R}{{\mathbb R}}
\usepackage {amsbsy}

\begin{document}
\maketitle
%\linenumbers

\begin{center}
Dedicated to Prof. Vincenzo Ciancio on his 70th birthday
\end{center}
%\linenumbers
%-------------------------------------------------------------------------------
\begin{abstract}
Materials with memory, namely those materials
whose mechanical and/or thermodynamical behaviour depends on 
time not only via the present time, but also 
through its past {\it history}, are considered.  Specifically, a three dimensional
viscoelastic body is studied.  Its mechanical behaviour is described via an
integro-differential equation, whose  kernel  represents the {\it relaxation modulus},
characteristic of the viscoelastic material under investigation. According to the
classical model, to guarantee the thermodynamical compatibility of the model itself, such
 a kernel satisfies regularity conditions which include the integrability 
of its time derivative. To adapt the model to a wider class of materials,  this condition is
relaxed; that is,  conversely to what is generally assumed, no integrability  condition 
is imposed on the time derivative of the relaxation modulus. Hence, 
 the case of  a relaxation modulus  which is unbounded  at the initial time $t=0$, is
 considered, so that
  a {\it singular kernel} integro-differential equation, is studied. In this framework, the
  existence of a weak solution is proved in the case of a three dimensional 
singular kernel initial boundary value problem.
\end{abstract}

 \bigskip

\noindent


%\linenumbers

\section{Introduction}
This note is concerned about viscoelasticity problems.  The interest on viscoelastic 
materials and their mechanical response to external actions is testified by a wide literature 
which goes from pioneering works due to Boltzmann \cite{Boltzmann} and Volterra 
\cite{Volterra} to a the theoretical study  developed to understand and model new 
materials also artificially devised. 
An overview on models which refer to materials 
requiring a {\it non-classical memory kernel} is provided in \cite{Intech2016} and in 
the references therein.  Specifically, the model adopted throughout is due to Giorgi and 
Morro    \cite{Giorgi-Morro92}, and, later, reconsidered by Gentili \cite{Gentili}. 
The crucial feature of all viscoelasticity models  is that the stress response at time $t$ 
linearly depends on the whole {\it past history} of the strain up to the present time $t$;
hence, an integral term appears in the governing equation. 
For sake of simplicity, 
the body is generally assumed to be homogeneous and isotropic: this assumption allows to 
consider that all the physically relevant quantities do not depend on the space variable. 
Currently, there is a growing interest in viscoelastic materials due to  new materials,
such as viscoelastic polymers and/or bio-materials. Thus, recent models are concerned about 
modelling living tissues \cite{Zhuravkov, arterie, Yue} via suitable viscoelasticity models; however, 
often, to provide a mathematical description which might fit with the behaviour in the case of new 
materials the classical regularity requirements imposed on the relaxation modulus are not appropriate.
In particular, less restrictive conditions correspond to relaxation moduli which are unbounded at the
initial time $t=0$ and, hence, correspond to a {\it singular kernel} model. Thus, Berti  \cite{Berti}, 
Grasselli and Lorenzi  \cite{Grasselli-Lorenzi91} and, more recently, Conti et al. \cite{Conti et al} 
are concerned about non-standard kernel models. Singular kernel problems, are widely investigated 
by many authors in various different applicative contexts 
\cite{CPV, Desch Grimmer, Enelund et al, Gentili, Hanyga, HanygaS, JW, Messaoudi, Miller-Feldstein, RHN}. Among the many, 
 the  book by Borcherdt   \cite{Borcherdt}  represents an overview on new materials 
 paying attention to applications of  viscoelastic models to seismic problems.
Fractional derivatives models are shown by   Fabrizio  \cite{Fabrizio2014} to represent,
in the case  of  a singular kernel, a possibility  to investigate  viscoelasticity  problems. 
 The interrelation between fractional derivatives and 
 viscoelasticity  \cite{Adolfsson et al, Hristov, Koeller, Mainardi-intro, Mainardi} seems to be promising also
  under the perspective of bio-materials   \cite{Deseri, arterie, Zhuravkov} or anisotropic homogeneous or 
  non-homogeneous materials \cite{Hilton}. Further to investigations which are concerned about the model 
  to describe a physical behaviour,
there are corresponding studies aiming to establish existence  and, possibly, also uniqueness of
solutions such as \cite{Berti, DIE2013, NONRWA2017}. 

The results here presented are part of a research project, the author is involved in,
which   concerns materials with memory, their 
behaviour as well as the study of related initial boundary value problems. 
Hence,  further to isothermal  viscoelasticity \cite{ACGM2014, DIE2013} 
also rigid heat conduction with memory \cite{SAM, CVV2013a} as well as, 
similarities between the two different cases under the analytical viewpoint \cite{JNMP2005, MECC2015}
are  investigated.
More recently,  magneto-viscoelasticity problems are studied in \cite{GVS2010, GVS2012, NONRWA2017}, 
motivated by new materials which are devised incorporating magnetically sensible
nanoparticles in a viscoelastic gel. 

The material is organised as follows. The opening Section 2 provides an overview 
on the classical model. Thus,  the strain tensor, the stress tensor and the
 relaxation modulus are introduced together with the functional requirements they 
are assumed to satisfy.  Then, the linear integro-differential 
equation which models the viscoelasticity problem is written. 
Finally, the classical Dirichlet problem studied by Dafermos \cite{Dafermos} is recalled.

In the next Section 3 the singular kernel problem under investigation is stated.
Then, on introduction of the integrated relaxation tensor, an equivalent formulation of 
the problem is obtained. An approximation strategy, as in \cite{DIE2013}, is devised  
  to construct a sequence 
of problems whose solution approximates the solution to the  problem under 
investigation. A Lemma which gives an estimate, crucial for subsequent results,  is 
proved.  

Section 4 contains the main existence result: the singular problem under investigation
is proved to admit solution. Specifically, via a suitable weak 
formulation of the problem, which takes into account the prescribed initial data and 
boundary conditions, a sequence of approximated solutions  
is proved to admit a limit which turns out to solve the singular problem.

The closing Section 5 is concerned about  some perspectives and open problems as well as 
connections with other works or related subjects. 

\section{A regular viscoelasticity problem}
 This Section is concerned about the introduction of the model of viscoelastic body; then,  
to  a regular viscoelasticity problem. specifically,  a $3$-dimensional {\it smooth} body
 whose  reference configuration is a smooth compact set $\Omega\subset \R^3$ is considered.
 Accordingly, the key features of the model of {\it viscoelastic body}, following 
 \cite{FabrizioMorro92, Giorgi-Morro92, Gentili},  are 
 briefly recalled. The material is assumed to be a {\it material with memory} to stress that its mechanical response 
 depends on time not only through the {\it present}  time  $t$ but also on the whole {\it past history}. Hence, when the viscoelastic 
 body is  assumed  homogeneous and isotropic is considered, so that the spatial dependence can be omitted in all the quantities of interest, that is, let $\mathbb{E}$ be the symmetric tensor
\begin{equation}
\displaystyle{\mathbb{E}:=\frac12\Big[\nabla{\bf u} +\nabla{\bf u}^T\Big]}
\end{equation}
 then 
\begin{equation}
\displaystyle{\mathbb{E} = \mathbb{E} \,(t)~~,~~\mathbb{T} = \mathbb{T} \,(t)~~,~~\mathbb{G} = \mathbb{G} \,(t)}
\end{equation}
 represent, in turn, the {\it strain tensor} \,$\mathbb{E} \in Sym$, the {\it stress tensor}\, $\mathbb{T} \in Sym$ and the {\it relaxation modulus} \,$\mathbb{G} \in Sym$. According to 
 \cite{Volterra, Gentili}, when  $\mathbb{G}_0:= \mathbb{G}(0)$ denotes the {\it instantaneous elastic modulus}, 
 the  following constitutive assumption links  strain and  stress tensors %via the relaxation modulus
        {\begin{equation}\label{2b}
        \displaystyle{\mathbb{T}\, (t) = \mathbb{G}_0
        \mathbb{E}(t) +\int_{0}^{\infty} { {\mathbb{G}}(\tau)~
        \dot{\mathbb{E}}^{t} \,(\tau) ~d \tau } ~,~\mathbb{E}^{t} \,(\tau):= \mathbb{E}(t-\tau)}
        \end{equation}}
where $\mathbb{E}^{t} $ is termed {{strain past history}}.
The latter, can be, equivalently, written as
\begin{equation}\label{2b2}
\displaystyle{\mathbb{T}\, (t) = \mathbb{G}_0
\mathbb{E}(t) +\int_{0}^{\infty} { \dot{\mathbb{G}}(\tau)~
\mathbb{E}^{t} \,(\tau) ~d \tau } }~.
\end{equation}
The relaxation modulus in  (\ref{2b}) and  (\ref{2b2}) satisfies the following regularity requirements
\begin{equation}\label{regularity}
{\mathbb{G}} \in L^1(\R^+)~~,~~ \dot{\mathbb{G}} \in L^1(\R^+, Lin(Sym))
\end{equation}
so that
\begin{equation}\label{regularG}
~~{\mathbb{G}}(t) = {\mathbb{G}}_0
+ \displaystyle{\int_{0}^{t} { \dot{\mathbb{G}}(s) ~d s}~,~{\mathbb{G}}(\infty)} =
 \displaystyle{\lim_{t\to \infty}{\mathbb{G}}(t)},
\end{equation}
where $\mathbb{G}(\infty)\in  Lin(Sym)$ is termed {\it equilibrium elastic modulus} \cite{FabrizioMorro92}. 
 Hence, the relaxation modulus ${\mathbb{G}}$ enjoys the {\it fading memory property},  that is 
 \begin{equation}\label{9}
\displaystyle \forall\varepsilon > 0\, \exists \, \tilde a = a \,(\varepsilon,
\mathbb{E}^t)\in\R^+ \textstyle{s.t.}~\forall a > \tilde a, 
\displaystyle{\left\vert\int_{0}^{\infty}\!\! { \dot {\mathbb{G}}(s+a)
\mathbb{E}^t(s) ~d s} \right\vert \! < \! \varepsilon }.
\end{equation}
Now, the integro-differential equation which gives the displacement 
 ${\bf u}: \Omega\times (0,T) \to \R^3$ in the case of a viscoelastic material, reads
 \begin{equation} \label{eq-visco}
\displaystyle { { \rho {\bf u}}_{tt} -{\rm div} \left(\mathbb{G}(0)\nabla{\bf u}+\int^t_0 ( \dot{\mathbb{G}}(t-\tau)\nabla{\bf u}(\tau) d\tau \right ) = {\bf f}},
\end{equation}
wherein the parameter $\rho \in\R^+$ can be, without loss of generality, assumed to be $1$, i.e. let $\rho=1$: this is the choice throughout.  
In particular, on introduction of the external force  ${\bf f}$ in which, further to an external force (optional), 
also the history of the material is included,  the following  initial boundary value problem, wherein the integral is posed in 
Volterra form 
 \begin{equation} \label{eqmve}
\left \{ \begin {array}{l} 
\displaystyle { {   {\bf u}}_{tt} -\mathbb{G}(0)\Delta{\bf u}+\int^t_0  \dot{\mathbb{G}}(t-\tau)\Delta{\bf u}(\tau) d\tau  = {\bf f}}\\ \\
{\bf u}(0)={\bf u}^0,\quad {{\bf u}}_t(0)={\bf u}^1 \,~  {\rm in}~~ \Omega\,;~~ {{\bf u}}=0 ~~~~\mbox {on}~~ \Sigma=\partial\Omega \times (0,T)
\end{array}\right.
\end{equation}
is considered. 
Under the classical regularity requirements (\ref{regularity}) an existence and uniqueness of 
a weak solution is proved by Dafermos \cite{Dafermos}. Crucial to establish the existence result is an
{\it apriori} estimate which relies on the regularity assumptions (\ref{regularity}) the
symmetric tensor $\mathbb{G}$ is  classically assumed to 
satisfy  \eqref{regularity}-\eqref{regularG}, which, 
%%%%%%%%%%%% SPOSTARE ?? DA QUI
in particular,  imply
 \begin{equation}\label{Gt}
\mathbb{G}\in L^1(0,T)\cap C^2(0,T) ,  %~~~~~~ \dot G \not\in L^1(0,T)~,
~~ \forall T\in \R~.
\end{equation}
%%%%%%%%%%%% SPOSTARE ?? FINO A QUI
In addition,
 (see, for instance 
 \cite{FabrizioMorro92, Gentili}): 
 % the relaxation function $G$, the kernel in the linear integro-differential equation, satisfies
   \begin{equation}\label{G}
    \mathbb{G}(t)>0,\qquad \dot{\mathbb{G}}(t)\le 0, \qquad \ddot{\mathbb{G}}(t)\ge 0,\qquad
    t\in(0,\infty),
    \end{equation}
that is,  the tensor's entries of $\mathbb{G}$, are such that, for any symmetric tensor 
$ e_{kl}$ 
\begin{eqnarray}\label{model_m}
 \begin{array}{ll}
 \bullet & {\mathbb{G}}_{klmn}={\mathbb{G}}_{mnkl}={\mathbb{G}}_{lkmn}  ~~ \\
 \bullet &{\mathbb{G}}_{klmn}\, e_{kl} e_{mn} \ge \beta\,  e_{kl} e_{kl}, \qquad
\beta\,\,>\,\,0,~e_{kl}= e_{lk} ~~.
~~~~~~~~~~~~~~~~~~~~~~~~~~
  \\
 \bullet &\dot{{\mathbb{G}}}_{klmn} e_{kl} e_{mn} \le 0 \\
 \bullet & \ddot{{\mathbb{G}}}_{klmn} e_{kl} e_{mn} \ge 0
\end{array}
\end{eqnarray}
Note that these sign conditions are crucial to prove Lemma 1, namely the {\it apriori} estimate on which the solution existence result is based.

{\section{Singular Memory Kernel Problem}}
 This Section is concerned about a singular viscoelasticity  problem which 
 represents %the $3$-dimensional analog of the regular problem mentioned in the previous Section. 
 a generalisation of the regular one presented in the previous Section. Specifically, a  $3$-dimensional 
 singular viscoelasticity problem is proved to admit a unique solution generalising the result obtained
  in \cite{DIE2013}, where the $1$-dimensional case is studied. 
An analogous result, see  \cite{GVS2012} for details, can be stated also in 
$3$-dimensional rigid thermodynamics.
  
 To take into account a wider class of materials, as specified for instance in \cite{MECC2015, Intech2016}, the regularity assumptions on $\mathbb{G}$ are relaxed. In particular the request 
 (\ref{regularity})$_2$ is removed, that is, we replace  (\ref{regularity}) with (\ref{Gt}).
% \begin{equation}\label{singularity}
%{\mathbb{G}} \in L^1(\R^+)~~. %~~ \dot{\mathbb{G}} \notin L^1(\R^+)~~.
%%~~ \mathbb{G} \in L^1(0,T),\forall T\in \R 
%\end{equation}
 Hence, the tensor $\mathbb{G}$ is unbounded at the origin and, therefore, the the integro-differential 
problem cannot be written under the form  (\ref{eq-visco})  where $\mathbb{G}(0)$, not defined, appears. 
To overcome this difficulty, following \cite{DIE2013}, observe that condition (\ref{Gt})
guarantees that the  
 {\it integrated relaxation tensor}  $\mathbb{K}$ can be defined via
 \begin{equation}\label{K}
\displaystyle \mathbb{K}(\xi):=\! \int_0^\xi \!\mathbb{G}(\tau) d\tau~~~~  \text{which implies}~~ ~~\mathbb{K}(0)=0~.
\end{equation}
Then,  the following integral problem
 \begin{equation}\label{Int-zero} 
\displaystyle{ {\rm P}: ~~{\bf u}({t})=\int_0^t \mathbb{K}(t-\tau) \Delta {\bf u}(\tau) d\tau + {\bf u}^1 t+ {\bf u}^0+\int_0^t d \tau\int_0^\tau {\bf f}(\xi) d\xi\,}\end{equation}
%where, respectively, ${\bf u}^0, {\bf u}^1, \f$ are the initial data and the external force in \eqrf{eqmve}, 
where, respectively, ${\bf u}^0, {\bf u}^1, {\bf f}$ denote the initial data and the external force, which, 
once again, %as  in \eqref{eqmve}, 
includes the past history, is well defined and represents an equivalent formulation of the i.b.v.p. \eqref{eqmve}.
Now, let  the {\it translated relaxation tensor}  be introduced via
\begin{equation}
\mathbb{G}^{\varepsilon}(\cdot):=\mathbb{G}({\varepsilon}+\cdot)~~,~~ \varepsilon >0~,
\end{equation}
which, recalling condition (\ref{Gt}), is well defined $\forall  \varepsilon >0$. 
Hence, if the initial time $t_0=\varepsilon$ is considered, the following integro-differential the problem P$_D^{\varepsilon}$ can be defined 
\begin{equation}\label{Geps} 
\displaystyle{{\rm P}_D^{\varepsilon}: ~~
{\bf u}^{\varepsilon}_{tt}= \mathbb{G}^{\varepsilon}(0) \, \Delta {\bf u}^{\varepsilon}+\int_0^t 
\dot{\mathbb{G}}^{\varepsilon}(t-\tau) \, \Delta {\bf u}^{\varepsilon}(\tau) \, d\tau +{\bf f}\, }~.
\end{equation}
Imposing on the latter  the initial and boundary conditions
\begin{equation} \label{ibcuG}
 {{\bf u}}^{\varepsilon}\vert_{t=0}={\bf u}^0(x),\quad {{\bf u}}^{\varepsilon}_t\vert_{t=0}
={{\bf u}}_1(x),
\quad {{\bf u}}_x^{\varepsilon}\vert_{\partial\Omega\times(0,T)}=0~~, ~~ t< T
\end{equation}
 a {regular}  problem, which can be regarded as an   approximation of the singular 
problem \eqref{Int-zero}, is obtained. The arbitrariness of $\varepsilon$, allows to look for a solution of 
the singular problem of interest according to the following steps.
\smallskip
\begin{itemize}
\item constructed a suitable sequence 
$\{{\rm P}^{{\varepsilon_h}}\}_{h\in \N}$ of {\it approximated regular problems} 
${\rm P}^{\varepsilon_h}$; 
\item find approximated solutions, denoted as ${{\bf u}}^{\varepsilon_h}$, the solution to  
%$P^{\varepsilon_h}$ in 
the {regular} problem P$_D^{\varepsilon}$  \eqref{Geps}-\eqref{ibcuG}; 
\item{show that the sequence of solutions admits a limit, i.e. $\displaystyle{\exists ~ {\bf u}:= \lim_{\varepsilon\to 0} {\bf u}^{\varepsilon}}$.}
%\item{recognize  {{uniqueness}} of  ${\bf u}$ solution to the  { {singular} } problem. }
\end{itemize}

{\subsection{Approximation Strategy %: Idea
}}

\noindent
The generic {Approximated Problem}  P$_D^{\varepsilon}$   \eqref{Geps}-\eqref{ibcuG},  fixed $\varepsilon > 0$, reads
 \begin{equation} \label{Peps}
{\rm P}_D^{\varepsilon}: ~\left \{ \begin {array}{l} 
\displaystyle {{\bf u}^{\varepsilon}_{tt}= \mathbb{G}({\varepsilon}) \, \Delta {\bf u}^{\varepsilon}+\int_0^t 
\dot{\mathbb{G}}({\varepsilon} +t-\tau) \, \Delta {\bf u}^{\varepsilon}(\tau) \, d\tau +{\bf f}\, }  \\ \\
%{\bf u}(0)={\bf u}^0,\quad \dot{{\bf u}}(0)={\bf u}^1, ~~
{\bf u}^{\varepsilon}(0)={\bf u}^0,\quad {{\bf u}^{\varepsilon}}_t(0)={\bf u}^1 \,~  {\rm in}~~ \Omega\,;~~ {{\bf u}^{\varepsilon}}=0 ~~~~\mbox {on}~~ \Sigma=\partial\Omega \times (0,T) ~~.
\end{array}\right.
\end{equation}
%where the choice $\varepsilon < 1$ is only due to the interest in the limit 
% when $\varepsilon \to 0$, can be written as 
The latter, according to \cite{Dafermos}, admits a unique solution, here denoted as 
${\bf u}^{\varepsilon}$: it is such that the following Lemma holds.
\medskip

\noindent{\bf Lemma 1} {\it Let ${\bf u}^{\varepsilon}$ %^0\in H^1_0(\Omega;\mathbb{R}^3)$,
%        ${\bf u}^1 \in L^2(\Omega;\mathbb{R}^3)$ and ${\bf f}
%        \in L^2(\mathcal{Q},\mathbb{R}^3)$ then we have
be, according to {\rm \cite{Dafermos}}, the solution to \eqref{Peps}, then the following equality holds
\begin{multline} \label{form1}
 \frac{1}{2}\frac{d}{dt}
\int_{\Omega}\mathbb{G}(t)\nabla{\bf u}^{\varepsilon} \cdot \nabla{\bf u}^{\varepsilon} d {\bf x} +\frac{1}{2} \frac{d}{dt}\int_{\Omega}\vert{\bf u}_{t}^{\varepsilon}\vert^2 d {\bf x}%\\ \\
-\frac{1}{2} \frac{d}{dt}\int_0^t ds \int_{\Omega}
\dot{\mathbb{G}}(s) \left|\nabla
{\bf u}^{\varepsilon}(t)-\nabla {\bf u}^{\varepsilon}(t-s)\right|^2
% \cdot  \left[\nabla {\bf u}^{\varepsilon}(t)-\nabla {\bf u}^{\varepsilon}(t-s) \right]
d {\bf x} =
\\ \\
= \int_{\Omega} {\bf f} \cdot {\bf u}_{t}^{\varepsilon}\, d {\bf x} +
 \frac{1}{2} \int_{\Omega} \dot{\mathbb{G}}(t)\nabla
%{\bf u}^{\varepsilon} \cdot \nabla {\bf u}^{\varepsilon} d {\bf x} -\hphantom{bcccccccccccaaaaaaaaaaa}\\
%\\
-\frac{1}{2} \int_0^t ds \int_{\Omega} \ddot{\mathbb{G}}(s) \left|\nabla
{\bf u}^{\varepsilon}(t)-\nabla {\bf u}^{\varepsilon}(t-s)\right|^2 
%\cdot  \left[\nabla {\bf u}^{\varepsilon}(t)-\nabla {\bf u}^{\varepsilon}(t-s) \right]
d {\bf x}~.
\end{multline}}

\medskip
\noindent{\bf Proof}. Consider  the integro-differential P$_D^{\varepsilon}$, subject to the assigned initial and boundary conditions, equation  (\ref{Peps})$_1$ can  be written in the equivalent form \eqref{eq-visco} as %, to equivalent  (\ref{Peps})
%  in the
%following equivalent form
\begin{equation} \label{eqmvepA1}
\\
\displaystyle{   {\bf u}_{tt}^{\varepsilon} -{\rm div} \left(\mathbb{G}(\varepsilon)\nabla{\bf u}^{\varepsilon}_{t}\int^t_0  \dot{\mathbb{G}}(\varepsilon+s)\left[\nabla{\bf u}^{\varepsilon}_{t}(t)-
 \nabla{\bf u}^{\varepsilon}_{t}(t-s)\right] ds \right)} %\\
= {\bf f}~.
\end{equation}
\\
The latter on multiplication  by ${\bf u}_{t}^{\varepsilon}$, followed by integration over $\Omega$, gives
\begin{equation} \label{form2}
\begin{array} {lll}
{\displaystyle
\frac{1}{2}\int_{\Omega}\mathbb{G}(t)
\nabla{\bf u}\cdot \nabla {\bf u}_{t} \,d {\bf x} +\frac{1}{2}
\frac{d}{dt}\int_{\Omega}\vert{\bf u}_{t}\vert^2 \,d {\bf x} }+\\
\\
+{\displaystyle \int_{\Omega} {{\bf u}_t}(t) \, d {\bf x}\, \int_0^t {\rm
div}\left[\dot{\mathbb{G}}(s) \,(\nabla {\bf u}(t)-\nabla {\bf u}(t-s))\right] ds }
 ={\displaystyle \int_{\Omega} {\bf f} \cdot {\bf u}_{t}\, d {\bf x} }~,
\end{array}
\end{equation}
wherein all  the indices $\varepsilon$ are omitted to simplify the notation. 
Then,  re-writting the first and the third terms in a more convenient form, \eqref{form2} reads
%\footnote{ recognize the time derivative 1st term, and by parts integration of the 3rd term}
%%%%% TORNA QUI !!!
\begin{equation} \label{form3}
\begin{array} {lll}
{\displaystyle
 \frac{1}{2}\frac{d}{dt}
\int_{\Omega}\mathbb{G}(t)\vert\nabla{\bf u} \vert^2
%\cdot \nabla {\bf u}
 \,d {\bf x} -\frac{1}{2}\int_{\Omega}\dot{\mathbb{G}}(t) 
\vert \nabla{\bf u} \vert^2 %\cdot \nabla{\bf u} 
\,d {\bf x} +  \frac{1}{2} \frac{d}{dt}\int_{\Omega}\vert{\bf u}_{t}\vert^2 d {\bf x}} -\\
\\
{\displaystyle  -\int_{\Omega} \int_0^t \dot{\mathbb{G}}(s) \,
\nabla {\bf u}_{t}(t)\cdot [\nabla {\bf u}(t)-\nabla {\bf u}(t-s)]
\, d {\bf x} ds
= \int_{\Omega} {\bf f}  \cdot {\bf u}_{t}\, d {\bf x} }~.
\end{array}
\end{equation}
%\medskip
Now, the last term on the right hand side can be written as follows:
\begin{equation}\label{useful}
\begin{array}{ccc}
\displaystyle{ -\int_{\Omega} \int_0^t \dot{\mathbb{G}}(s) \, \nabla
{\bf u}_{t}(t)\cdot [\nabla {\bf u}(t)-\nabla {\bf u}(t-s)] \, d {\bf x} ds=}\\ \\
 = \displaystyle{-\frac{1}{2} \frac{d}{dt}\int_0^t ds \int_{\Omega}
\dot{\mathbb{G}}(s) |\nabla {\bf u}(t)-\nabla {\bf u}(t-s)|^2 
%\cdot [\nabla {\bf u}(t)-\nabla {\bf u}(t-s)] d {\bf x} 
+}
\\ \\
\displaystyle{ +\frac{1}{2}\int_{\Omega} \int_0^t \ddot{\mathbb{G}}(s) |\nabla
{\bf u}(t)-\nabla {\bf u}(t-s)|^2 %\cdot [\nabla {\bf u}(t)-\nabla {\bf u}(t-s)] \, 
d {\bf x} ds~,}
\end{array}
\end{equation}
which, substituted into (\ref{form3}), proves Lemma 1. Here, for sake of completeness, detailed
computations which prove the identity \eqref{useful} are reported. 

\begin{multline*}
-\int_{\Omega} \int_0^t \dot{\mathbb{G}}(s) \, \nabla
{\bf u}_{t}(t)\cdot [\nabla {\bf u}(t)-\nabla {\bf u}(t-s)] \, d {\bf x} ds=\\
= -\frac{1}{2} \frac{d}{dt}\int_0^t ds \int_{\Omega}
\dot{\mathbb{G}}(s)\vert\nabla {\bf u}(t)-\nabla {\bf u}(t-s)\vert^2 
%\cdot [\nabla {\bf u}(t)-\nabla {\bf u}(t-s)] 
d {\bf x} +
\\
+\frac{1}{2} \int_{\Omega} \dot{\mathbb{G}}(s) 
\vert\nabla {\bf u}(t)-\nabla {\bf u}(t-s)\vert^2
%[\nabla {\bf u}(t)-\nabla {\bf u}(0)] \cdot [\nabla {\bf u}(t)-\nabla {\bf u}(0)]  
d {\bf x}  \,-
\\
\hphantom{bbbbbbbbbbbb}-\int_{\Omega} \int_0^t \dot{\mathbb{G}}(s) \nabla {\bf u}_{t}(t-s)\cdot
[\nabla {\bf u}(t)-\nabla {\bf u}(t-s)] \, d {\bf x} ds=
\\
%\end{multline*}
%\begin{multline*}
 = -\frac{1}{2} \frac{d}{dt}\int_0^t ds \int_{\Omega}
\dot{\mathbb{G}}(s)|\nabla {\bf u}(t)-\nabla {\bf u}(t-s)|^2 
%\cdot[\nabla {\bf u}(t)-\nabla {\bf u}(t-s)] d {\bf x} 
+
\\
\end{multline*}
\begin{multline*}
+\frac{1}{2} \int_{\Omega} \dot{\mathbb{G}}(s) |\nabla {\bf u}(t)-\nabla
{\bf u}(0)|^2 %\cdot [\nabla {\bf u}(t)-\nabla {\bf u}(0)] 
d {\bf x}  \,+
\\
\hphantom{bbbbbbbbbbbbbbbbbb}+\int_{\Omega} \int_0^t \dot{\mathbb{G}}(s) \,|\nabla {\bf u}(t)-\nabla
{\bf u}(t-s)|^2 %\cdot \frac{d}{ds}\nabla {\bf u}(t-s)
\, d {\bf x} ds=
\\
 = -\frac{1}{2} \frac{d}{dt}\int_0^t ds \int_{\Omega}
\dot{\mathbb{G}}(s) \vert\nabla {\bf u}(t)-\nabla {\bf u}(t-s)\vert^2 
%\cdot [\nabla {\bf u}(t)-\nabla {\bf u}(t-s)] 
d {\bf x} +
\\
\hphantom{bbbbbbbbbbbbbbbbb}+\frac{1}{2} \int_{\Omega} \dot{\mathbb{G}}(s) [\nabla {\bf u}(t)-\nabla
{\bf u}(0)] \cdot [\nabla {\bf u}(t)-\nabla {\bf u}(0)] d {\bf x}  \,-
\\
\hphantom{bbbbbbbb}-\int_{\Omega} \int_0^t \dot{\mathbb{G}}(s) 
[\nabla {\bf u}(t)-\nabla{\bf u}(t-s)] \cdot \frac{d}{ds}[\nabla {\bf u}(t)-\nabla {\bf u}(t-s)]\, d {\bf x}
ds=
\\
 = -\frac{1}{2} \frac{d}{dt}\int_0^t ds \int_{\Omega}
\dot{\mathbb{G}}(s) \vert\nabla {\bf u}(t)-\nabla {\bf u}(t-s)\vert^2
%\cdot [\nabla {\bf u}(t)-\nabla {\bf u}(t-s)]
 d {\bf x} +
\\
 +\frac{1}{2}\int_{\Omega} \int_0^t \ddot{\mathbb{G}}(s) 
 \vert\nabla {\bf u}(t)-\nabla {\bf u}(t-s)\vert^2
% [\nabla{\bf u}(t)-\nabla {\bf u}(t-s)] \cdot [\nabla {\bf u}(t)-\nabla {\bf u}(t-s)]
  \, d {\bf x} ds~.
\end{multline*}%\vert\nabla {\bf u}(t)-\nabla {\bf u}(t-s)\vert^2
 This completes the proof of \eqref{useful} and, hence, of Lemma 1.

 \hfill $\square$

\medskip
\noindent Then, the further estimate can be proved.
%%%%%%%%%%%%%%%% DA DCDS FINE- da EECT INIZIO 

\medskip

\noindent{\bf Lemma 2} {\it Let ${\bf u}^{\varepsilon}$ denote a solution  to \eqref{Peps}, then 
the following estimate holds
\begin{equation}\label{2.10}
\displaystyle{ {1\over{2}} \int_{\Omega}  \vert \nabla {u}\vert^2\, d {\bf x} + 
{1\over{2}}
\int_{\Omega} \vert{u}_t\vert^2\, d {\bf x} \le \gamma e^T
C({{\bf f}}) } 
\end{equation}
wherein $\gamma= \max\{(\vert\mathbb{G}(T+1)\vert)^{-1}, 1\}.$}

\medskip
\noindent{\bf Proof}. 

\bigskip\noindent
Consider (\ref{form1}) and integrate it w.r.to time, over the time interval $(0,t)$,  since ${\bf u}^{\varepsilon}$
is a solution of the integro-differential problem   P$_D^{\varepsilon}$ \eqref{Peps}, it satisfies also the assigned
initial and boundary conditions; hence
\begin{equation}\label{ineq1}\begin{array}{cl@{\hspace{0.5ex}}c@{\hspace
{1.0ex}}l}
 \displaystyle{  \frac{1}{2}
\int_{\Omega}\mathbb{G}(t)\nabla{\bf u}^{\varepsilon} \cdot \nabla{\bf u}^{\varepsilon} d {\bf x} +\frac{1}{2}\int_{\Omega}\vert{\bf u}_{t}^{\varepsilon}\vert^2 d {\bf x}}\le\qquad\qquad\qquad\qquad\qquad
\\ \\
\qquad \displaystyle{\le  \int_0^t  \int_{\Omega} {\bf f} \cdot {\bf u}_{t}^{\varepsilon}\, d {\bf x} +
 \frac{1}{2}  \int_0^t \int_{\Omega} \dot{\mathbb{G}}(t)\nabla
{\bf u}^{\varepsilon}(0) \cdot \nabla {\bf u}^{\varepsilon}(0) d {\bf x} 
+ {1\over{2}}  \int_0^t \int_{\Omega} 
\vert{{\bf u}_t({\bf x},0)}\vert^2\, d {\bf x}}\,
\\ \\
\qquad\qquad\qquad\qquad \displaystyle{\le  \int_0^t \int_{\Omega} {\bf f} \cdot {\bf u}_{t}^{\varepsilon}\, d {\bf x} 
+ {1\over{2}}  \int_0^t \int_{\Omega} 
\vert{{\bf u}_1}\vert^2\, d {\bf x}}\,.
\end{array}\end{equation}
If, in addition,   homogeneous  initial  conditions are imposed, then, the last inequality reduces to
%        
%        \`E VERO ??
%        here considered, reads: 
\begin{equation}\label{ineq1C}
 \displaystyle{  \frac{1}{2}
\int_{\Omega}\mathbb{G}(t)\nabla{\bf u}^{\varepsilon} \cdot \nabla{\bf u}^{\varepsilon} d {\bf x} +\frac{1}{2}\int_{\Omega}\vert{\bf u}_{t}^{\varepsilon}\vert^2 d {\bf x} \le  \int_0^t  \int_{\Omega} {\bf f} \cdot {\bf u}_{t}^{\varepsilon}\, d {\bf x} }
\end{equation}
Then,  it follows, 
\begin{equation}\label{est}
{1\over{2}} \int_{\Omega}\mathbb{G}(t)\nabla{\bf u}^{\varepsilon} \cdot \nabla{\bf u}^{\varepsilon} d {\bf x} + 
{1\over{2}} \int_{\Omega} \vert{{\bf u}}_t\vert^2\, d {\bf x} - \int_0^t \int_{\Omega}
\vert{{\bf u}}_t\vert^2\, d {\bf x} ds\le C({{\bf f}}) ~,
\end{equation}
when ${\bf u}_1=0$, otherwise, in the latter and also in the following estimates,  $C({{\bf f}})$
 should be replaced by $C({{\bf f}}, {\bf u}_1)$ .
%Note that the first term in the latter is related to the free energy 
Gronwall's lemma applied to \eqref{est} gives
\begin{equation}
\displaystyle{{1\over{2}}\int_{\Omega}\mathbb{G}(t)\nabla{\bf u}^{\varepsilon} \cdot \nabla{\bf u}^{\varepsilon} 
d {\bf x}+  {1\over{2}} \int_{\Omega} \vert{{\bf u}}_t\vert^2\, d {\bf x} \le e^T C({{\bf f}}) }~,
\end{equation}
which, when the sign conditions  \eqref{G} and \eqref{model_m} are recalled,  $\vert\mathbb{G}(t+\varepsilon)\vert\geq \vert\mathbb{G}(T+1)\vert$, and therefore
\begin{equation}\label{2.10}
\displaystyle{ {1\over{2}} \int_{\Omega}  \vert \nabla {u}\vert^2\, d {\bf x} + 
{1\over{2}}
\int_{\Omega} \vert{u}_t\vert^2\, d {\bf x} \le \gamma e^T
C({{\bf f}})~~, ~~ \gamma:= \max\{(\vert\mathbb{G}(T+1)\vert)^{-1}, 1\}. } 
\end{equation}

\hfill$\square$

\bigskip

{\section{Solution Existence}}
This Section is concerned about the existence result: the main 
one obtained in this note. Specifically, the existence of a weak solution is
stated in Theorem 1, together with a suitable weak formulation of the problem; 
then, after the proof's sketch, a Lemma is stated and proved.
The Section closes with the proof of the Theorem.

\medskip 
Let $\left\{{\rm P}_D^{{\varepsilon}_h}\right\}_{h\in\N}$ denote a sequence of 
approximated problems ${\rm P}_D^{\varepsilon}$,   \eqref{Geps}, 
consider the corresponding sequence of admitted solutions $\left\{{\bf u}^{{\varepsilon}_h}\right\}_{h\in\N}$,
each of which satisfies also the assigned initial and boundary conditions \eqref{ibcuG}.
Theorem 1 shows that given a solution sequence, it 
admits a limit which represents a solution to the singular problem under investigation.
Specifically,  the following Theorem can be stated.
\medskip
 
 { {\rm\bf Theorem 1}}~~\label{Existence}
{Given ${\bf u}^{\varepsilon}$ solution to the integral problem P$_I^{\varepsilon}$
 \begin{equation}\label{pb2-eps} 
\hskip-3em \displaystyle{{\rm P}_I^{\varepsilon}:  {\bf u}^{\varepsilon}({t})=\int_0^t \mathbb{K}^{\varepsilon}(t-\tau) \Delta {\bf u}^{\varepsilon}(\tau) d\tau + {\bf u}_1 t+ {\bf u}_0+\int_0^t d \tau\int_0^\tau {\bf f}(\xi) d\xi\, },
\end{equation} 

\begin{equation}\label{lim} 
 \displaystyle{\exists \,\, {\bf u}(t)=\lim_{\varepsilon \to 0} {{\bf u}^{\varepsilon}({t})}\,
~~ \text{in} ~~ {\bf L}^2({\mathcal{Q}})~,~{\mathcal{Q}}=\Omega\times(0,T).}
\end{equation} }
\medskip

%  {\bf {Proof:}} 
\noindent  The outline of the proof, based on the equivalence between the integral formulation \eqref{pb2-eps} 
  and the integro-differential one  \eqref{Geps}-\eqref{ibcuG},  is based on the following steps:
 \medskip 
\begin{itemize}
\item{consider the weak formulation of the problem; }
 \item{consider separately the terms which do not dependend on ${\varepsilon}$;}
 \item{consider the terms where ${\bf u}^{\varepsilon}~ \&~ \mathbb{K}^{\varepsilon}$ appear;}
 \item{apply the convergence result in Lemma 2.} % prove convergence via Lebesque's Theorem 
\end{itemize} 

Previous to the proof of the Theorem, the weak formulation of the problem and a
 Lemma are crucial to prove the existence theorem. First of all, on introduction of 
 the following test functions 
\begin{equation}\label{test-fun}
\displaystyle{
\mathbf{v}\equiv(v_1,v_2,v_3),~~ v_i\in H_0^1({\mathcal{Q}})~~,~~ {\mathcal{Q}}=\Omega\times(0,T), ~~\text{s.t.}~~~{\mathbf{v}=\mathbf{0}}~~\text{on}~~~\partial\Omega~,%~~ {\mathcal{Q}}=\Omega\times(0,T)~, 
}\end{equation}
the weak formulation of the problem can be constructed asserting that, given 
${\bf u}^{{\varepsilon_h}}$ it represents a {\it weak} solution to \eqref{pb2-eps} whenever
\begin{equation}\label{weak-eps}
 \displaystyle{\int_{\mathcal{Q}}  \mathbf{v} \cdot{\bf u}^{\varepsilon_h}({t}) dtd {\bf x}=
 \int_{\mathcal{Q}}  \mathbf{v} \cdot\left[\int_0^t \mathbb{K}^{\varepsilon}(t-\tau) \Delta {\bf u}^{\varepsilon_h}(\tau) d\tau + {\bf u}_1 t+ {\bf u}_0+\int_0^t d \tau\int_0^\tau {\bf f}(\xi) d\xi\,\right]  dtd {\bf x} },
\end{equation}
for all test functions ${\mathbf{v}}$ in \eqref{test-fun}. 
 
 \noindent
Consider, now, the following Lemma, crucial  to prove the existence theorem.
 \medskip
 
 \noindent
{\bf{Lemma 2} }
{\it %The integral problem  {\rm(\ref{pb2-eps})}, where 
Given $\mathbb{K}(\varepsilon)$, which, according to \eqref{K},  is 
 \begin{equation}\label{Keps}
\displaystyle \mathbb{K}^{\varepsilon_h}(\xi)=\! \int_0^\xi \!\mathbb{G}({\varepsilon_h}+\tau) d\tau~~~~ ,% \text{and hence}~~ 
~~\mathbb{K}^{\varepsilon_h}(0)=0~~~, \forall {\varepsilon_h}~,
\end{equation}
and  any test function  $\mathbf{v}$, defined  in \eqref{test-fun},
then it follows that 
\begin{equation}\label{l1} 
 \displaystyle{\lim_{{\varepsilon_h}\to 0} \int_0^T \!\!  \!\! \int_{\Omega} \Delta \mathbf{v} \cdot 
\int_0^t \left[\mathbb{K}^
{{\varepsilon_h}}(s)- \mathbb{K}(s)\right] {\bf u}^{{\varepsilon_h}} (t-s) d s  d {\bf x} dt  =0~}
\end{equation}}
%\end{lemma}

\noindent	{\bf{Proof}} {  Note that,  $\forall ( {\bf x},t)\in \Omega\times(0,T)$ 
	
\medskip	
\begin{equation}
\displaystyle{ \vert \mathbf{u}\vert \le C \vert \Omega \vert, }~~~~~~~~
\displaystyle{ \vert \Delta\mathbf{v}\vert \le M, }
\end{equation}
furthermore
\begin{equation}
\displaystyle{ \left\vert\mathbb{K}^{\varepsilon_h}(s)-\mathbb{K}(s)\right\vert =
\left\vert\mathbb{K}({\varepsilon_h}+s)- \mathbb{K}(s)\right\vert= 
\int_s^{{\varepsilon_h}+s} \mathbb{G}(\tau) d \tau}
\end{equation}
hence, since $\mathbb{G}\in L^1(0,T)$,  Lebesgue's Theorem implies the limit convergence.	

\hfill  $\square$ }%\eeproof}}

Now, the Theorem can be easily proved. 

\noindent	{\bf{Proof of Theorem 1}}{  
To start with, consider the approximated problems in their formulation \eqref{Peps}, 
the estimate  (\ref{2.10})  guarantees there is a subsequence 
$\{ {\varepsilon_h}\}, h\in \N$ such that there exists a convergent subsequence 
of solutions $\{ {\bf u}^{\varepsilon_h}\}$  
\begin{equation}\label{24}
{\bf u}^{{\varepsilon_h}}\longrightarrow~~ {\bf u}~~ \textrm{weakly ~in} ~{\bf H}^1(0,T, 
H^1_0(\Omega))~~ \textrm{as} ~ {{\varepsilon_h} \to 0};
\end{equation}
%%%%%
%%%%%                            OPPURE METTO H BOLD !!!!!         ?????????????
%%%%%
\begin{equation}\label{25}
{\bf u}^{{\varepsilon_h}}\longrightarrow~~ {\bf u}~~ \textrm{strongly ~in} ~ {\bf L}^2(\Omega
\times(0,T))  
~~ \textrm{as} ~ {{\varepsilon_h} \to 0};
\end{equation}
hence 
\begin{equation}\label{cclim} 
 {\displaystyle{\exists \,\, {\bf u}(t)=\lim_{{\varepsilon_h} \to 0} {{\bf u}^{{\varepsilon_h}}}({t})}\,
~~ \textrm{in} ~~ {\bf L}^2(\Omega\times(0,T)),}
\end{equation}
where ${\bf u}^{{\varepsilon_h}}$ is solution to the problem \eqref{Geps}-\eqref{ibcuG}. 
This convergence result allows to prove Theorem 1. Consider the weak formulation of 
the integral problems P$_I^{{\varepsilon_h}}$ expressed in (\ref{pb2-eps}). 
%Accordingly,  let us introduce the {\it test  functions} $\mathbf{v}$, which depend on both the time and space variables and 
%satisfy homogeneous boundary conditions at t the smooth {\it boundary} 
%$\partial\Omega$ of   $\Omega\subset\R^3$. 
%\begin{equation}\label{icphi}
%\mathbf{v}\equiv(v_1,_2,v_3), v_i \in C^\infty(\Omega\times(0,T))~, i=1,2,3
%~~ \textstyle{s.t.}~~ \mathbf{v} \vert_{\partial\Omega}= \mathbf{0} ~~~ \forall t\in  (0,T)~,
%\end{equation}
Scalar  multiplication of (\ref{pb2-eps}) by $\mathbf{v}$ followed by integration over 
$\Omega\times(0,T)$, delivers \eqref{weak-eps}.
The following integral  
\begin{equation}\label{t1}  
\displaystyle{ 
\int_0^T \!\!  \!\! \int_{\Omega} \!\!  \mathbf{v} \cdot\left\{
 {\bf u}_1 t+ {\bf u}_0+\int_0^t  \!\!d \tau \!\!\int_0^\tau {\bf f}(\xi) d\xi\,\right\} d {\bf x} dt .}
\end{equation}
collects all the terms which appear in \eqref{weak-eps},  and do not dependent on $\varepsilon_h$. 
Such terms, conversely, depend on  the history of the material as well as on
the initial conditions, all of them assumed to be regular. Hence, since the 
integration domain $ {\mathcal{Q}}=\Omega\times(0,T)$ is bounded, it follows that also
 the integral over $\Omega\times(0,T)$, in (\ref{t1}), is bounded. Furthermore, since such
 integral does not depend  on ${{\varepsilon_h}}$, then it is unchanged in the limit 
 ${{\varepsilon_h}}\to 0$.  As a consequence,
 the only term which remains to consider, in the limit, is 
\begin{equation}\label{pbint-eps} 
 \displaystyle{ 
  \int_0^T \!\!  \!\!  \int_{\Omega} \mathbf{v} \cdot \left[  \int_0^t \mathbb{K}^{{\varepsilon_h}}(t-\tau) 
 \Delta u^{{\varepsilon_h}} (\tau) d \tau   \right] d {\bf x} dt}~.
\end{equation}
Let $\tau=t-s$, then %, %via change of variables, it follows
\begin{equation}\label{pbint-eps3} 
 \displaystyle{  \int_0^t \mathbb{K}^{{\varepsilon_h}}(t-\tau)\Delta u^{{\varepsilon_h}}(\tau) d\tau =  
\int_0^t \mathbb{K}^{{\varepsilon_h}}(s) \Delta  u^{{\varepsilon_h}} (t-s) d s} ~.
\end{equation}
On use of the homogeneous boundary conditions imposed on  the test functions 
$ \mathbf{v}$, on integration over $ \Omega$, two times, we obtain
\begin{equation}\label{pbint-eps2} 
 \displaystyle{
 \int_0^T\!\!  \!\!  \int_{\Omega} \left[\mathbf{v} \cdot \int_0^t \mathbb{K}^{\varepsilon_h}(t-\tau) 
 \Delta u^{\varepsilon_h} (\tau) d \tau\right]   d {\bf x} dt  =    \int_0^T \!\!  \!\! \int_{\Omega} \left[\Delta \mathbf{v} \cdot
   \int_0^t \mathbb{K}^{{\varepsilon_h}} (t-\tau) u^{{\varepsilon_h}} (\tau) d\tau \right]  \, d {\bf x} dt
~.}
\end{equation}
Then, recalling (\ref{pbint-eps3}),  the r.h.s. of the latter can be equivalently 
 expressed as
 \begin{equation}\label{pbint-eps4}  
\begin{array}{cl@{\hspace{0.5ex}}c@{\hspace{1.0ex}}l}   
 \displaystyle{ 
  \int_0^T \!\!  \!\! \int_{\Omega} \Delta \mathbf{v} \cdot \int_0^t \mathbb{K}^{{\varepsilon_h}}(t-\tau) 
u^{{\varepsilon_h}}(\tau) d
\tau   d {\bf x} dt =\qquad\qquad\qquad\qquad}\\ \\ \displaystyle{\qquad
 =  \int_0^T  \int_{\Omega} \Delta \mathbf{v} \cdot \int_0^t \left[\mathbb{K}^{{\varepsilon_h}}(s)- \mathbb{K}(s)
\right] u^
{{\varepsilon_h}}(t-s) d s  d {\bf x} dt  +}\\ \\ \displaystyle{\qquad\qquad\qquad\qquad\qquad+ \int_0^T  \int_{\Omega} \Delta \mathbf{v} \cdot \int_0^t  
\mathbb{K}(s) u^{{\varepsilon_h}}(t-s) 
d s  d {\bf x} dt }~.
\end{array}\end{equation}
where the term $$\displaystyle{ \int_0^T  \int_{\Omega} \Delta \mathbf{v} \cdot \int_0^t  
\mathbb{K}(s) u^{{\varepsilon_h}}(t-s)  d s  d {\bf x} dt }$$%
is added and subtracted. Then, the Theorem is proved via combination of 
 (\ref{24}) -- (\ref{25})  with  Lemma 2.}

\hfill  $\square$

\section{Conclusions and Perspectives}
The result presented shows that solution existence holds also when the requirements on
the regularity of the relaxation modulus ${\mathbb{G}}$ are weaker than the classical ones. Namely, 
${\mathbb{G}} \in L^1$ but $\dot{\mathbb{G}} \notin L^1$. the achieved result generalises the previous 
ones. In particular, in  \cite{DIE2013}, the existence and uniqueness of an initial boundary value 
problem in the case of a $1-$dimensional viscoelastic body with a singular kernel. Notably, as a consequence
of the mathematical analogy between the models which describe isothermal viscoelasticity, on one side, and
rigid thermodynamics with memory, on the other one, singular kernel problems  in the two different
kind of materials  can be compared  \cite{MECC2015}. Thus, an existence result    \cite{CVV2013a}
can be proved in the case of a singular kernel problem in rigid thermodynamics. 
Furthermore, new materials are also obtained inserting nano-particles which are  magnetically active, in a
viscoelastic gel; in this way, materials which can be termed magneto-viscoelastic ones are obtained.
Magneto-viscoelasticity problems are considered  in \cite{GVS2010, GVS2012}  where the coupling of
the two different effects: viscoelastic response of the material and magnetisation is considered.
Singular kernel problems in magneto-viscoelastic materials are considered in \cite{NONRWA2017}.
New perspective investigations are concerned about the possibility to model various different concomitant 
effects such as introduce also thermal  effects in the case of the viscoelastic body, or, instead of magnetic effects,
consider more general electro-magnetic ones, to mention only those effects which seems suitable not
only to be modelled, but also to be analytically studied in a near future.

\end{document}